\documentclass[preprint,12pt]{elsarticle}
\usepackage{cite}
\usepackage{booktabs}
\usepackage{multicol}
\usepackage{subcaption}
\usepackage{rotating}

\usepackage{amssymb}
\journal{Journal of Quantitative Spectroscopy and Radiative Transfer}

\begin{document}

\begin{frontmatter}


\title{Assessing light pollution in vast areas: zenith sky brightness maps of Catalonia}



\author{Hector Linares $^{1,2,3}$, Eduard Masana $^{3}$, Salvador J Ribas $^{4}$, Manuel García-Gil$^{5,6}$, Martin Aubé$^{1,2,7}$, Alejandro Sánchez de Miguel$^{8,9,10}$, Alexandre Simoneau$^{1,2}$\\}

\address{$^{1}$ D\'epartement de g\'eomatique appliqu\'ee, Universit\'e de Sherbrooke\\
$^{2}$ D\'epartement de physique, C\'egep de Sherbrooke\\
$^{3}$ Institut de Ci\`{e}ncies del Cosmos (ICC-UB-IEEC), Barcelona, Spain\\
$^{4}$ Parc Astron\`{o}mic Montsec - Ferrocarrils de la Generalitat de Catalunya,  \`{A}ger, Spain\\
$^{5}$ Departament d’Acci\'{o} Clim\`{a}tica, Alimentaci\'{o} i Agenda Rural, Generalitat de Catalunya\\
$^{6}$ Departament d'Enginyeria de Projectes i de la Construcci\'{o}, Universitat Polit\`{e}cnica de Catalunya, BarcelonaTech, Spain\\
$^{7}$ Physics Department, Bishop's University\\
$^{8}$ Depto. F\'{i}sica de la Tierra y Astrof\'{i}sica. Instituto de F\'{i}sica de Part\'{i}culas y del Cosmos (IPARCOS), Universidad
Complutense, Madrid, Spain\\
$^{9}$ Instituto de Astrof\'{i}sica de Andaluc\'{i}a, Glorieta de la Astronom\'{i}a, Granada, Spain\\
$^{10}$ Environment and Sustainability Institute, University of Exeter, Penryn, Cornwall, U.K.\\}

\begin{abstract}
Zenith sky brightness maps in the V and B bands of the region of Catalonia are presented in this paper. For creating them we have used the light pollution numerical model Illumina v2. The maps have a sampling of 5x5 km for the whole region with an improved resolution of 1x1 km for one of the provinces within Catalonia, Tarragona. Before creating the final maps, the methodology was tested successfully by comparing the computed values to measurements in nineteen different locations spread out throughout the territory. The resulting maps have been compared to the zenith sky brightness world atlas and also to Sky Quality Meter (SQM) dynamic measurements. When comparing to measurements we found small differences mainly due to mismatching in the location of the points studied, and also due to differences in the natural sky brightness and atmospheric content. In the comparison to the world atlas some differences were expected as we are taking into account the blocking effect of topography and obstacles, and also due to a more precise light sources characterization. The results of this work confirm the conclusion found in other studies that the minimum sampling for studying sky brightness fine details is of 1x1 km. However, a sampling of 5x5 km is interesting when studying general trends, mainly for vast areas, due to the reduction of the time required to create the maps.
\end{abstract}


\end{frontmatter}


\section{Introduction}
\label{section:introduction}
Assessing the quality of the night sky for vast areas is a big challenge. The rapid changes of sky brightness with distance and atmospheric conditions make it difficult to use measurements at a set of individual locations for that purpose, however, some attempts have been made \citep{linares_thesis}.

Our purpose is to define a methodology for assessing large areas using a light pollution model instead of measurements. In this sense, Falchi et al \citep{falchiatlas} created a zenith sky brightness (ZSB) world atlas map combining data from the satellite sensor VIIRS-DNB and Garstang's light pollution model \citep{Garstang,garstang1986model}. This atlas gives information in the V band with a spatial resolution of approximately 1 km$^2$. In fact, the work was an update from Cinzano et al \citep{cinzano2, cinzano2001} that instead of VIIRS-DNB data, which was not available yet, used DMSP-OLS satellite data information. The final product of \citep{cinzano2} was a ZSB map of Europe in B and V bands, and in \citep{cinzano2001} the map was expanded to a world atlas map but only for the V band. To reduce the computational time needed to create the atlas in \citep{falchiatlas}, several assumptions were made: same atmospheric conditions for every location, constant city emission function, constant spectrum of artificial lights and ignoring orographic screening effects. The results from that work had a huge impact in the light pollution community and even beyond the scientific scope. It allowed to check pollution levels with ease and to compare pollution among countries and regions. It revealed that most of the world population lives under polluted skies.

An alternative approach for creating a zenith sky brightness world map was introduced by S\'anchez de Miguel \citep{sanchezdemiguelVIIRS}. Their methodology did not rely on any light pollution model, but was based on the idea that the radiance received (from outside city borders) by the VIIRS-DNB is strongly correlated to the radiance received on the ground. %This idea comes from two assumptions. First, that molecular scattering can be modeled as Rayleigh scattering that has back-forward and azimuthal symmetry. And second, that aerosol scattering can be modeled as Mie scattering. Although Mie scattering has a peak in the forward direction, when studying ZSB outside cities, most of the artificial light comes from low elevation angles (horizontally directed light) that is scattered by the atmosphere close to zenith. Mie scattering azimuthal symmetry predicts that in this case similar amounts of light are scattered upwards and downwards.

For this project, we decided using the Illumina radiative transfer model \citep{aube2005,aube2018new, canarias} for two main reasons. First, because Illumina takes into account variables ignored by the methods previously explained, such as orographic screening and heterogeneous city emission functions (spectral and optical properties). Second, because the methodology defined in this paper for creating the ZSB maps can be used in the future with different brightness indicators, with changing lighting scenarios and any spectral band within the working range of the model.

\section{Methods}

\subsection{Illumina}

Illumina is a sophisticated light propagation numerical model able to simulate with precision the human induced sky radiance. Its scheme is similar to ray-tracing software: a statistically selected set of photons is thrown from luminaries to the atmosphere and interactions with the ground, molecules and aerosols are computed along the light paths towards a simulated observer \citep{canarias}.

It accounts for first and second order scattering without limitation in direction or distance. This numerical approach is very time consuming due mostly to second order scattering computation. Although the model allows to run cases only accounting for single scattering contribution, it is not interesting for the purposes of this study as the second order scattering relative contribution rises with distance from city limits and may contribute up to 66\% of the total zenith radiance for remote sites \citep{aube2007light}. In this sense, Kocifaj \citep{kocifaj2018} also concluded that higher orders than first order scattering are not negligible for distances superior to 30 km for blue light and 60 km for red light. Therefore, it is crucial for cases as the one presented in this paper with many locations far from cities.

Illumina can use, and even combine, different sources of information for computing the radiance emitted by light sources. In our study we have used satellite data from the VIIRS-DNB for that purpose. The model is able to compensate for the lack of information in the blue part of the spectrum from the VIIRS-DNB by using the spectral information of light sources and the VIIRS-DNB spectral sensitivity \citep{aube2018new}. 

%Illumina, among other options, uses satellite data from the VIIRS-DNB to compute the radiance emitted by light sources. The model is able to compensate for the lack of information in the blue part of the spectrum from the VIIRS-DNB by using the spectral information of light sources and the VIIRS-DNB spectral sensitivity \citep{aube2018new}. 

Topographic information is also gathered using satellite data. The model includes obstacles, such as buildings and trees, to define each source emission function. Angular and spectral emission functions are estimated by the combination of luminaries and lamps within the city.

The model assumes that the atmospheric content is constant for the whole geographic domain when estimating the sky brightness over a location. It also assumes a plane-parallel atmosphere for dealing with the interaction between photons and atmospheric particles.

%The main features of the model are:

%\begin{enumerate}
%\item First order and second order molecular and aerosol scattering. It depends on: the transmittance of the atmosphere in every path, defined as the fraction of incident electromagnetic radiation that is transmitted by the atmosphere; the scattering probability; and the scattering phase function (see \ref{atm}).
%\item Vertical variations in atmospheric optical properties (see \ref{atm}).
%\item Satellite data for source definition. Non uniform ground reflectance values and non-uniform distribution of light fixture luminosity (see \ref{source}).
%\item Heterogeneous characterisation of lamps. It allows to combine different spectra and angular patterns within the same town, and define differently light sources within an experiment (see \ref{source}).
%\item Topographic information to determine differences in altitude between observer and light sources and also for determining the orographic blocking effects (see \ref{scenario1}).
%\end{enumerate} 
Illumina produces two outputs:
\begin{enumerate}
    \item The artificial radiance received in each location, direction and wavelength chosen.
    \item Radiance contribution maps. For each combination of spectral window and direction studied, Illumina provides a map showing the relative contribution of each surface source to the total radiance received.
\end{enumerate}

\subsection{ISS derived inventories}
\label{iss}
\begin{figure}[h!]
\begin{center}
\includegraphics[width=12cm]{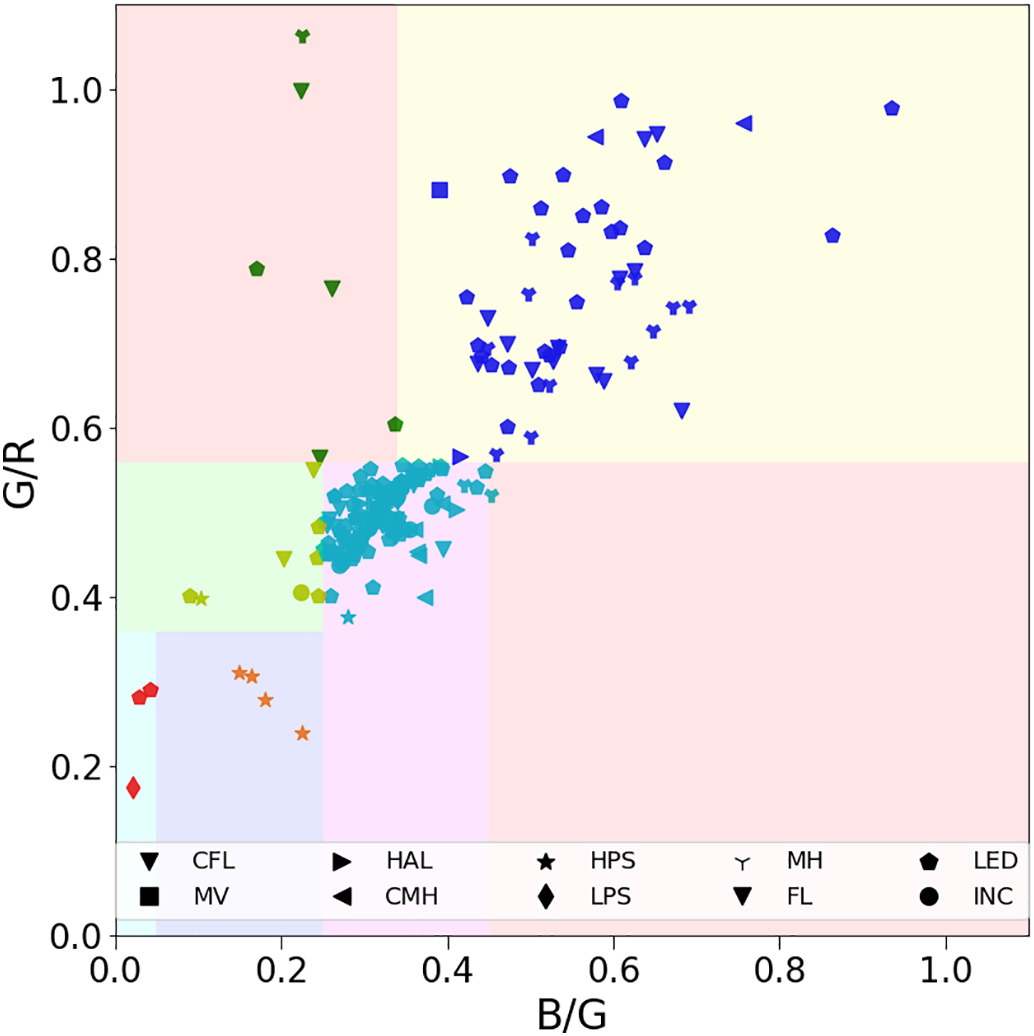}
\caption{DSLR channel ratios of the most common light fixtures technologies: Compact Fluorescent (CFL), Mercury Vapor (MV), Halogens (HAL), Ceramic Metal Halide (CMH), High Pressure Sodium (HPS), Low Pressure Sodium (LPS), Metal Halide (MH), Fluorescent (FL), Light Emitting Diode (LED), Incandescent (INC). From \citep{sanchez2019}.}
\label{tech_ratios}
\end{center}
\end{figure}

The spectral definition of light sources was mainly done following the methodology introduced by S\'anchez de Miguel \citep{sanchez2019}\citep{SANCHEZDEMIGUEL2021112611} that uses calibrated images taken from the International Space Station (ISS) to derive the spectrum of emitted light. The strategy followed is:
\begin{itemize}
    \item Estimating the radiance ratios (defined as \textit{colours} in astronomy) between the three bands (RGB) of the DSLR cameras that the most used lighting technologies produce (see figure \ref{tech_ratios}).
    \item Obtaining the radiance ratios or \textit{colours} of the pixels from the images taken with DSLR cameras on board of the ISS.
    \item Comparing both ratios, the ones from the images and the one from the most used lighting technologies, and assigning a dominating technology to every pixel of the image taken from the ISS.
    \item Estimating the lighting system of a source (city or town) as a mix of different technologies weighted by radiance taking into account all the pixels from the image that lie within the city/town borders. 
\end{itemize}

\subsubsection{Map characteristics}

Illumina v2 has been used to create sky brightness maps of Catalonia (Spain) and Tarragona with the following characteristics: they show zenith sky brightness, have a spatial resolution of 5x5 km for Catalonia (1310 locations) and 1x1 km for Tarragona (6300 locations), cover the B and V bands of the Johnson system and study the current lighting system. We chose this configuration to maximize the computational resources that we have access to. The methodology is the same when working with other indicators or higher spatial resolution.

The same atmospheric characterization was used for all the locations. As aerosol content at night is quite difficult to get, it is typically estimated using a worldwide network called Aerosol Robotic Network (AERONET) that tracks is optical depth during daytime. The typical values from the two AERONET stations within the territory (Montsec and Barcelona) for April, May and June of the years 2019 and 2020 were used: relative humidity 70\%, clear sky, Angstrom coefficient of 1.0, and aerosol optical depth of 0.1 at 500 nm.  

\begin{figure}[h!]
\begin{center}
\includegraphics[width=15cm]{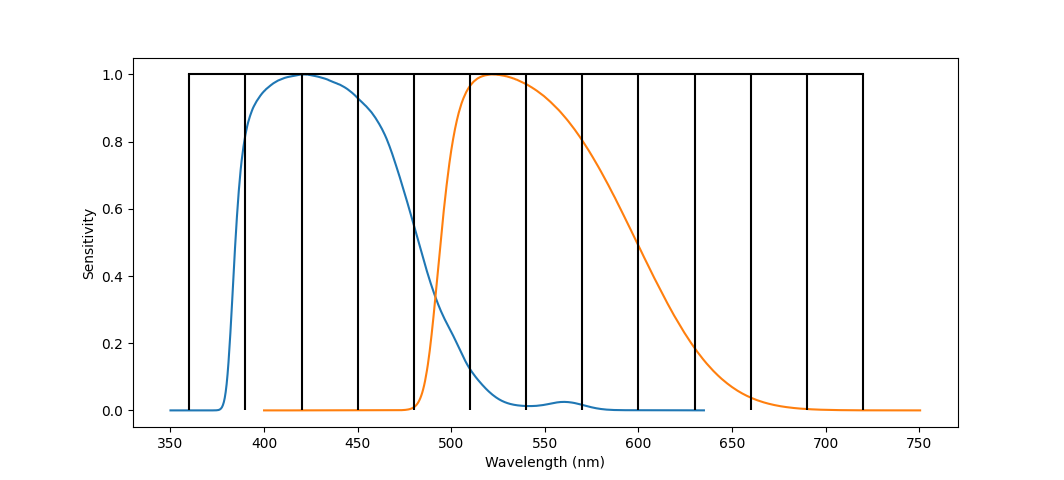}
\caption{Filter B (blue), filter V (orange), and the 12 spectral windows of 30 nm computed with Illumina.}
\label{windows}
\end{center}
\end{figure}

The spectral range studied goes from 360 nm to 720 nm to fully cover the B and V bands of the Johnson-Cousins system (see figure \ref{windows}). The bands have been chosen because most of the artificial light is emitted within their range, and also for being able to compare the results to previous studies \citep{falchiatlas} (see section \ref{comparison_falchi1}). We divided this range in 12 rectangular windows of 30 nm that allow to estimate the values obtained in both bands (see figure \ref{windows}).

Illumina provides only artificial radiance so we added the natural brightness using a reference value \citep{walker87} of 22.7 mag/arcsec$^{2}$ in B and 21.8 mag/arcsec$^{2}$ in V filters. Bear in mind that in the absence of artificial brightness those would be the values present in the map.

\begin{figure}[h!]
\begin{center}
\includegraphics[width=8cm]{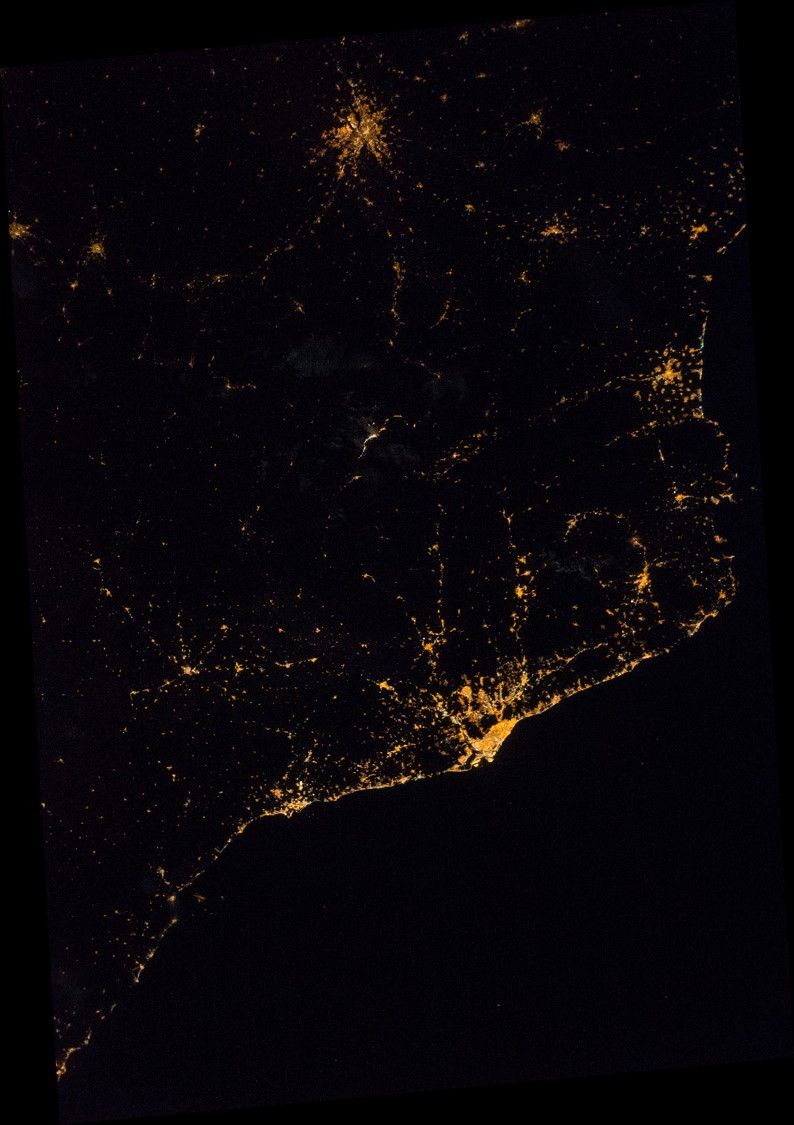}
\caption{Picture taken from the ISS centered in Catalonia. NASA Photo ID: ISS052-E-31962. Source: Earth Science and Remote Sensing Unit, NASA Johnson Space Center.}
\label{ISSimage}
\end{center}
\end{figure}

The image ISS052-E-31962 from the Earth Science and Remote Sensing Unit, NASA Johnson Space Center taken the 8th of April 2017 (see figure \ref{ISSimage}) was used for estimating light inventories through the method explained in section \ref{iss}. They are shown in table \ref{inventory_zbm} (labelled as Info Source 1).

The VIIRS image used is the monthly average composite from the month of April 2018 provided by the "Earth Observation group" \footnote{https://eogdata.mines.edu/products/vnl/} \citep{earth_observation_group}. The month was chosen to match the same month when the ISS image was taken, and the year to match the same year when the measurements used for validation and comparions of results (see section \ref{validation} and \ref{sqm_measurements} respectively) were taken.

Inventories of small cities and towns are difficult to estimate with the same method because the signal to noise ratio is very small. For these cases, information provided by the \textit{Departament d’Acci\'{o} Clim\`{a}tica, Alimentaci\'{o} i Agenda Rural, Generalitat de Catalunya} was used for creating a general inventory applicable to all small sources (General zone in table \ref{inventory_zbm}).

\begin{table}
    \begin{center}
    \begin{tabular}{|p{3cm}|c|c|c|c|c|c|c|}
    \hline
    Light source  & HPS & 3LED & 4LED & MH & MV  & FLC & Info source\\
    \hline
    Lleida & 23 & 5 & 51 & 14 & 1 & 6 & 1\\
    Balaguer & 76 &14 & 7 & 2 & 0 &1 &1\\
    Tremp & 89 & 5 & 4 & 1 & 0 & 0 &1\\
    Vall\`{e}s & 53 & 13 & 19 & 6 & 2 & 7 &1\\
    Barcelona MA & 61& 15& 12& 4& 1& 5 &1\\
    Vilafranca del P. & 57& 14& 16& 6& 1& 7 &1\\
    Figueres & 50&13&25& 7& 1& 4 &1\\
    Girona & 66& 16& 8& 3& 1& 6 &1\\
    Igualada & 72& 13& 10& 3&0& 2 &1\\
    Manresa & 56& 12& 21& 6& 1& 5 &1\\
    TGN petrochemical plant & 35& 9& 14& 9& 6& 27 &1\\
    Reus & 59& 14& 18& 5& 1& 4 &1\\
    Tarragona MA & 31& 7& 37& 12& 2& 10 &1\\ 
    Salou & 47& 11& 9& 7& 4& 21 &1\\
    Vic & 69&17&9&3&0& 2 &1\\
    Andorra & 13& 3& 61& 16& 1& 5 &1\\
    Barbastro & 61& 13& 19& 5&0& 2 &1\\
    Bin\`{e}far & 44&8&33& 9& 1& 5 &1\\
    Fraga & 47& 13& 21& 7& 2& 9 &1\\
    Monz\'on & 53& 8& 28& 8& 1& 2 &1\\
    Perpignan & 56& 16& 16& 5& 1& 6 &1\\
    Toulouse & 54& 18& 17& 5& 1& 4 &1\\
    Vinar\'os & 21& 4& 43& 14&3& 15 &1\\
    Pen\'iscola& 21&4& 43& 14&3&15 &1\\
    Benicarl\'o&21&4&43&14&3&15 &1\\
    General zone (all other) & 85&0 & 15& 0 & 0 & 0 &2\\
    \hline
    \end{tabular}
    \caption{Percentage of the radiance emitted by lighting technology of each source. HPS: high pressure sodium. 3LED: LED 3000 K CCT. 4LED: LED 4000 K CCT. MH: metal halide.  MV: mercury vapour. FLC: fluorescent. Information source: 1 corresponds to the methodology from \citep{sanchez2019}\citep{SANCHEZDEMIGUEL2021112611} applied to the image ISS052-E-31962 from the Earth Science and Remote Sensing Unit, NASA Johnson Space Center; 2 corresponds to public inventories provided by the Departament d’Acci\'{o} Clim\`{a}tica, Alimentaci\'{o} i Agenda Rural, Generalitat de Catalunya.}
    \label{inventory_zbm}
   
    \end{center}
\end{table}

We compared the public inventories information provided by some city councils to the total estimated inventories of the same sources using the ISS image. We found that some cases are very similar, such as Balaguer, Tremp and Sort. But others, as in the city of Lleida both inventories are quite different. The public inventory is dominated by HPS technology (74\% HPS, 14\% LED 4000K, 12\% Metal halide) but the method from \citep{sanchez2019} and \citep{SANCHEZDEMIGUEL2021112611} shows that it is dominated by LED 4000 K CCT instead. This contradiction could be explained for two reasons: first, the inventory information that we had access to could not be precise/complete enough; second, the private lighting constitutes a big percentage of the light emitted by the city (the images taken from the ISS reveal spectrum information from public and private lighting). Detailed information from more cities is needed to reach a conclusion, but if the second hypothesis is confirmed it would mean that private lighting has a non-negligible impact on the emission spectrum of a city and should be taken into account.

Regarding urban architecture, sources have been divided in three different categories: cities over 50 000 inhabitants, towns between 1000 and 50 000, and towns with less than 1000 inhabitants (see table \ref{urban_params}). The numbers shown in the table have been derived from detailed inventories of lamps of some representative sources, such as Lleida, Balaguer and Tremp, as well as an exhaustive inspection of architectural features using virtual representations of some of the sources. The numbers derived were already used in \citep{linares2020}.

\begin{table}[h]
\begin{center}
\begin{tabular}{|p{5.5cm}|c|c|c|}
\hline
Source type  & LFH (m) & HDO (m) & OH (m)\\
\hline
Big cities ($>$50k inhabitants)  & 7 & 6 & 10\\
Towns ($>$ 1000 inhabitants)  & 6 & 5 & 7\\
Towns ($<$ 1000 inhabitants) & 5 & 4 & 5\\
\hline
\end{tabular}
\caption{Urban parameters used as inputs for the maps. LFH: light fixture height. HDO: horizontal distance to obstacles. OH: obstacles' height. }
\label{urban_params}
\end{center}
\end{table}

\section{Validation}
\label{validation}
As stated previously, Illumina is very time consuming. As a matter of fact, the 5x5 km map of Catalonia took close to 50 000 CPU hours and the 1x1 km map of Tarragona close to 200 000 CPU hours.

\begin{figure}[h!]
\begin{center}
\includegraphics[width=14cm]{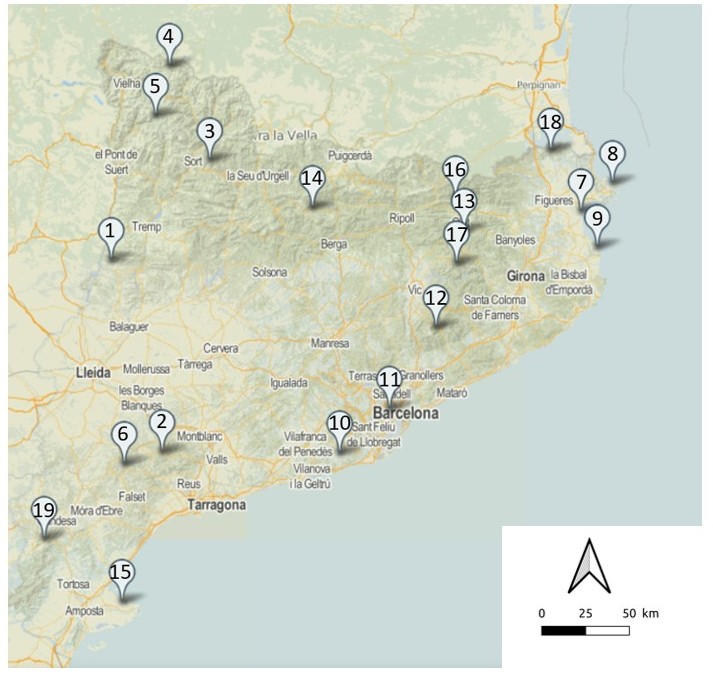}
\caption{Test locations where the ZBM was estimated and compared against measurements.}
\label{mapaloczbm}
\end{center}
\end{figure}

Before launching such a big job it was worth testing the methodology defined by running a small case. We computed the ZSB of 19 points spread within Catalonia (see figure \ref{mapaloczbm}) and compared the results to the measured reference value in each location. The locations were chosen with the aim to represent most of the territory. We also tried to select points with very different characteristics: some of them are close to the coastal line, others are in mountain regions, some are close to big cities, others are placed in very low population areas, etc.

The computed ZSB values of the 19 points are shown in table \ref{tablezbm} alongside the available measured ZSB values. For the measurements we used SQM, SQC and ASTMON data (see \citep{hanel2017measuring,Cinzano_sqm} for information about the instrumentation). 

The best option would be comparing results to ASTMON as it provides B and V band information, but only four locations had been measured with this device: OAdM, Prades, Montsant and Estany Llong/Aig\"uestortes. The best alternative to ASTMON are DSLR images processed by SQC. They provide ZSB in the V band but not in the B band. There are four locations measured with this method \citep{linares_thesis}: OAdM, Prades, Orri and Montgarri. Some of them were measured with the Milky Way close to zenith or abnormal absorption conditions, we acknowledged qualitatively the associated brightening effect when doing the validation. 

\begin{figure}[h!]
\begin{center}
\includegraphics[width=14cm]{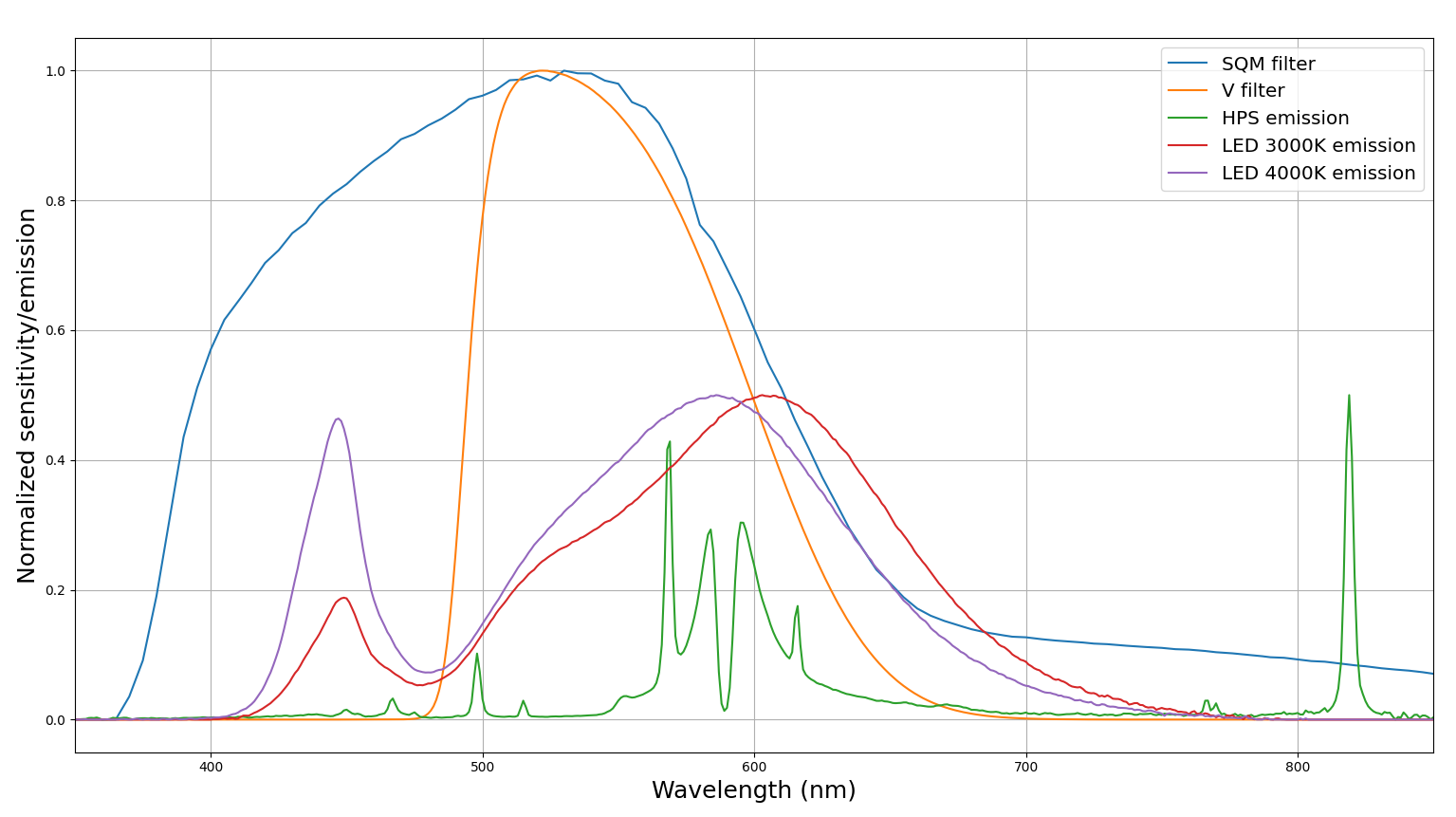}
\caption{Sensitivity of the V and SQM filters alongside the three most used technology lamps used in Catalonia. Sensitivity curves are normalized to a maximum of 1, and emission functions to a maximum of 0.5. Blue: SQM. Orange: V Johnson. Purple: LED 4000K. Red: LED 3000K. Green: HPS. }
\label{curves}
\end{center}
\end{figure}

\begin{figure}[h!]
\begin{center}
\includegraphics[width=14.5cm]{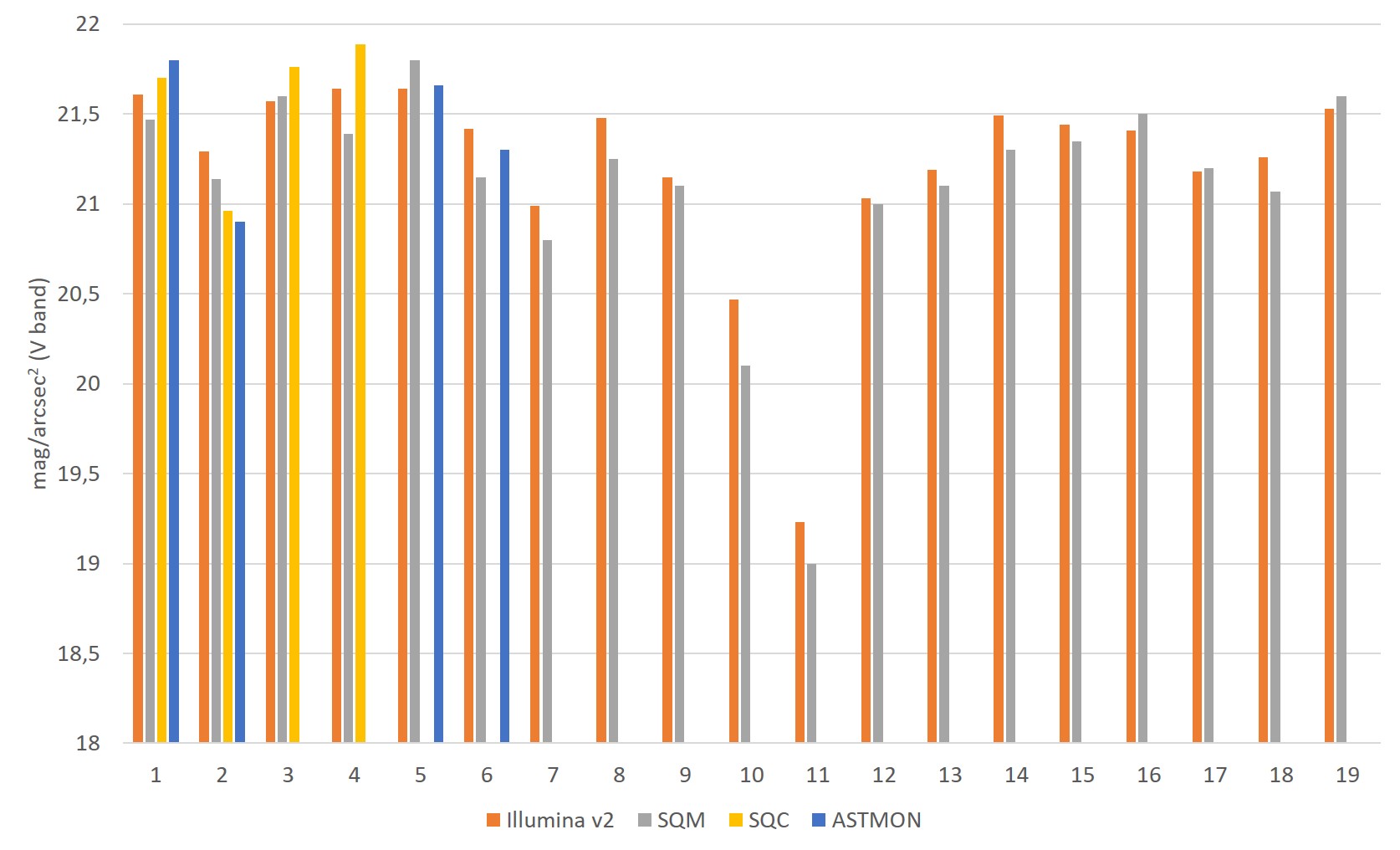}
\caption{Zenith sky brightness values of the 19 test locations. Reference number according to table \ref{tablezbm}.}
\label{chartzbm}
\end{center}
\end{figure}

The last option is using SQM values \footnote{The TESS \citep{tess} photometers were not used for the validation as there were only a few at the moment and they presented very similar problems to SQMs.}. The model used was the SQM-LU and the value used as reference is the average of six measurements taken within a minute every 10 seconds in the same location. Although SQM values were used as the main validation reference, it is not the best tool for two main reasons. First, SQM-LU measures the brightness of the sky in a wider region of the sky (approximately 20 degrees FWHM), normally the zenith is the darkest (if there is no Milky Way in that direction), thus, by measuring a larger area the measurement is normally slightly brighter than calculated for the zenith and other devices with narrower field of view. The second reason why the comparison with SQM is not ideal is because its filter is similar but not the same as the V band of the Johnson system (see figure \ref{curves}). Most typical lights present in the territory have peaks out of the range of the V but within the SQM sensitivity (see figure \ref{curves}), leading to brighter measurements in the SQM than in V. The difference then is expected to be bigger as more polluted is the sky (as artificial light dominates the sky brightness). As a matter of fact, this trend is observed in figure \ref{chartzbm} (it is not a linear trend as there are other variables involved): in dark locations differences are typically small (even a few locations show darker SQM values), on the other hand the three brightest locations (7, 10 and 11) show some of the biggest differences (being SQM brighter). A numerical study of this phenomenon is added as an appendix. 

\begin{sidewaystable}
\begin{center}
\begin{tabular}{|c|c|c|c|c|c|c|c|}
\hline
 &Location &  Sim. in V Band &  V ASTMON & V SQC & SQM & Sim. B & B ASTMON\\
\hline
1 & OAdM  & 21.61 & 21.8 & 21.70 & 21.47  & 22.63 &  22.8 \\
2 & Prades & 21.19 & 20.9 & 20.96 & 21.14 & 22.47 & 21.8 \\ 
3 & Orri & 21.57 & - & 21.76 & 21.60 & - & -\\
4 & Montgarri & 21.64 & - & 21.89 & 21.39 & - & - \\
5 & Aig\"uestortes  & 21.64 & 21.7 & - & 21.80 & 22.63 & 22.7 \\
6 & Montsant  & 21.42 & 21.3 & - & 21.15 & 22.54 & 22.5\\ 
7 & Aiguamolls Empord\`{a} & 20.99 & - & - & 20.80 &- &- \\
8 & Cap de Creus   & 21.48 & - & - & 21.25 &- &- \\
9 & Mass\'is Montgr\'i & 21.15 & - & - & 21.10 &- &-\\ 
10 & Garraf  & 20.47& - & - & 20.10  &- &-\\
11 & Collserola  & 19.23 & - & - &  19.00 &- &-\\
12 & Montseny   & 21.03 & - & - & 21.00  & - &- \\
13 & Santa Pau  & 21.19 & - & - & 21.10 & - & - \\ 
14 & Saldes   & 21.49 & - & - & 21.30 & - & -\\
15 & Deltebre  & 21.44 & - & - & 21.35 & - &- \\
16 & Vall d'en Bac  & 21.41 & - & - & 21.50 &- &- \\
17 & Rupit  & 21.18 & - & - & 21.20 & - & -\\
18 & Requesens  & 21.26 & - & - & 21.07 &- &- \\
19 & Els Ports & 21.53 & - & - & 21.60 & - &- \\
\hline
\end{tabular}
\caption{Comparison between measurements and computed zenith brightness. Measurements values are shown alongside the instrumentation used: SQM, SQC or ASTMON.}
\label{tablezbm}
\end{center}
\end{sidewaystable}

ZSB computed values agree well with measurements (see table \ref{tablezbm} and figure \ref{chartzbm}). There is no pattern in the sense of computed values being systematically brighter or darker than measured, with the exception of the SQM that is typically brighter for the reasons commented above. In table \ref{statistics} the mean differences and the standard deviation for each method of measurements compared to the simulated values are shown in mag/arcsec\textsuperscript{2}.

\begin{table}[h]
\begin{center}
\begin{tabular}{|p{5.5cm}|c|c|c|}
\hline
  & Mean difference & Standard dev.\\
\hline
Model V vs Astmon V  &  0.04 & 0.18\\
Model V vs SQC V  & -0.075 & 0.20\\
Model V vs SQM & 0.11 & 0.14\\
Model B vs Astmon B & 0.12 & 0.35 \\
\hline
\end{tabular}
\caption{Statistics of the comparison between simulated values and measured ones.}
\label{statistics}
\end{center}
\end{table}

The difference between the different measurements methods and the computed values are typically within 0.2 mag/arcsec\textsuperscript{2}. Taking into account that during prior campaigns \citep{linares_thesis} it has been measured that different atmospheric conditions may produce differences up to 0.3 mag/arcsec\textsuperscript{2} (same location, both moonless and cloudless, and far for the Milky Way direction), and that the intrinsic error of the measurements method is of the order of magnitude of 0.1 mag/arcsec\textsuperscript{2} \citep{kyba2011a}, we concluded that the methodology and definition of sources were validated, and we proceeded to create the ZSB maps in V and B bands.

\section{Results}
\label{section:results}
\subsection{General remarks}
\subsubsection{Catalonia maps (5x5 km)}
The core computation of the maps presented in this section have been carried out in the supercomputer Mammouth II from Calcul Québec and Compute Canada. The computation have been divided in 1310 processes, each process corresponds to the full calculus of one point (located in the middle of the cell in the map), meaning 12 independent jobs corresponding to the 12 wavelength bands in which the wavelength range is divided. The computational time for each job is not constant as it depends on the specific characteristics of each location in terms of spectrum and intensity of the sources around. The job mean time has been of approximately 3.2 hours, resulting in a ratio of 0.65 h/km\textsuperscript{2} due to each point is the reference for an area of 5x5 km. To comply with the limitations of the supercomputer processes were computed in packages of 200 that ran in parallel.

\begin{figure}[h!]
\begin{center}
\includegraphics[width=13cm]{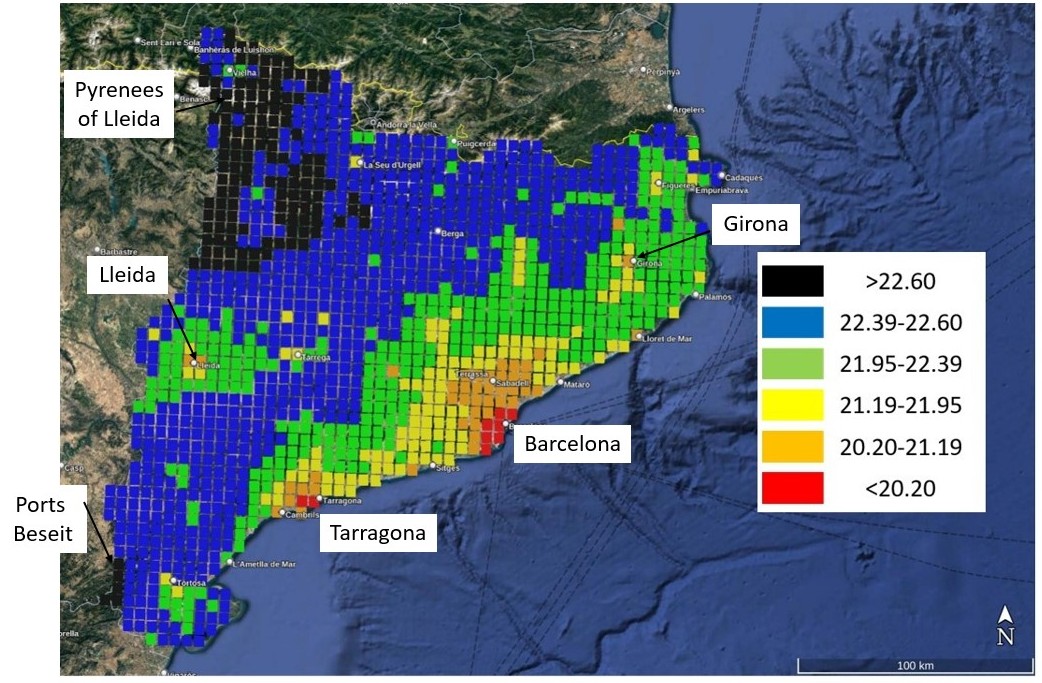}
\caption{Zenith sky brightness map of Catalonia in the B band.}
\label{ZSBM_B}
\end{center}
\end{figure}

\begin{figure}[h!]
\begin{center}
\includegraphics[width=13cm]{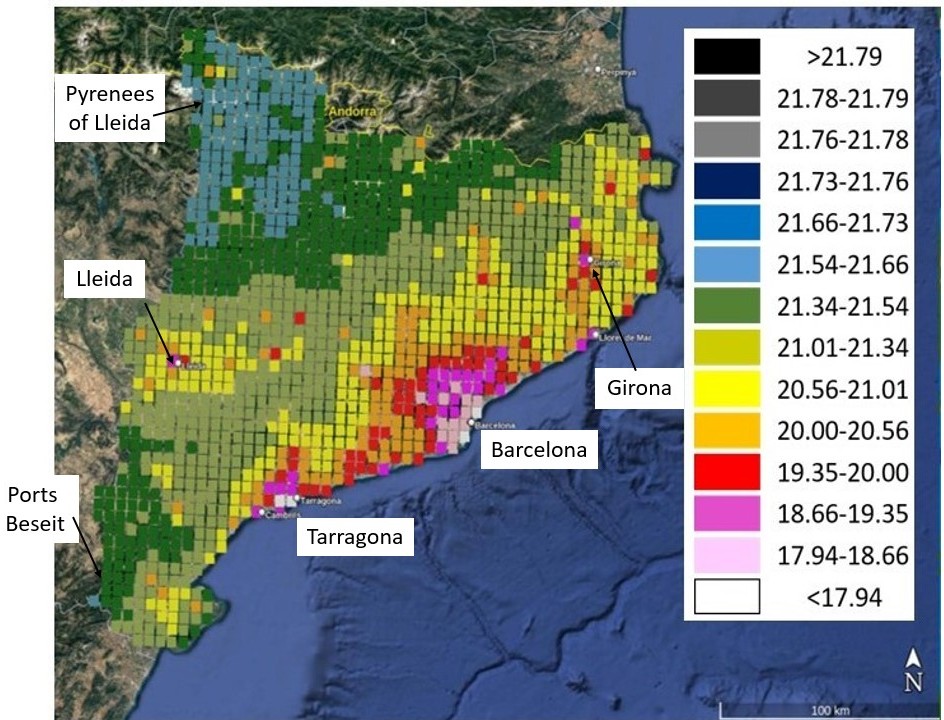}
\caption{Zenith sky brightness map of Catalonia in the V band.}
\label{ZSBM_V}
\end{center}
\end{figure}

The results for the B and V bands are displayed respectively in figure \ref{ZSBM_B} and \ref{ZSBM_V}, the color scale is different in each one for visual purposes. As stated previously, we have established reference values for each band in order to be able to analyse and compare the results: 21.8 mag/arcsec\textsuperscript{2} for the V band and 22.7 mag/arcsec\textsuperscript{2} for the B band \citep{walker87}.

Both maps show that there are two main focus of light pollution in Catalonia. The biggest one is Barcelona metropolitan area and the second one Tarragona metropolitan area. In the V band the two influence areas are almost fused together. The most polluted skies apart from the ones already mentioned are over the middle-size cities of Lleida and Girona.

The maps also coincide in displaying a gradient of darkness from the coastal line towards the inland. The darkest areas are in Pyrenees of Lleida (NW) and towards the Els Ports de Beseit NP (SW).

Regarding middle and small sized towns the sampling is not good enough for inferring any conclusion. If the sampling location lays within a small town it creates the false illusion of a bright area, as it covers 5x5 km, as a matter of fact there are some isolated cells with bright values in both maps. On the other hand, small but bright sources (as energy plants or skiing resorts) can be masked if the sampled locations are far away , i.e. two kilometers. This conclusion agrees well with \citep{Barasampling} where it is stated that "about one measurement per square kilometre could be sufficient for determining the artificial zenithal NSB at any point of a region to within 0.1 magV arcsec–2 (in the root-mean-square sense) of its true value in the Johnson–Cousins V band". The 1x1 km sampling map of Tarragona confirms it by providing more accurate information of the situations explained above.

\subsubsection{Tarragona map (1x1 km)}

The core computation of the maps presented in this section have been carried out in the supercomputer Pirineus II of the Consorci de Serveis Universitaris de Catalunya (CSUC). The computation have been divided in 6300 processes, each process corresponds to the full calculus of one point, meaning 12 independent jobs corresponding to the 12 wavelength bands in which the wavelength range is divided. The computational time of each job is not constant as it depends on the specific characteristics of each location in terms of spectrum and intensity of the sources around. The job mean time has been of approximately 2.6 hours, resulting in a ratio of 31.2 h/km\textsuperscript{2} as in this case  each point is the reference for an area of 1x1 km. To comply with the limitations of the supercomputer processes were computed in packages of 400 that ran in parallel.

\begin{figure}[h!]
\begin{center}
\includegraphics[width=11cm]{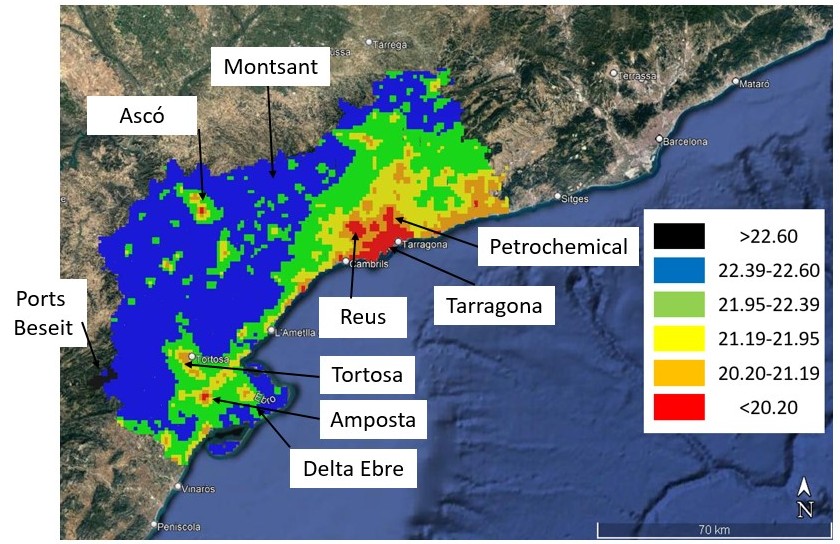}
\caption{ZSB maps in B band of Tarragona.}
\label{tarragona_B}
\end{center}
\end{figure}

\begin{figure}[h!]
\begin{center}
\includegraphics[width=11cm]{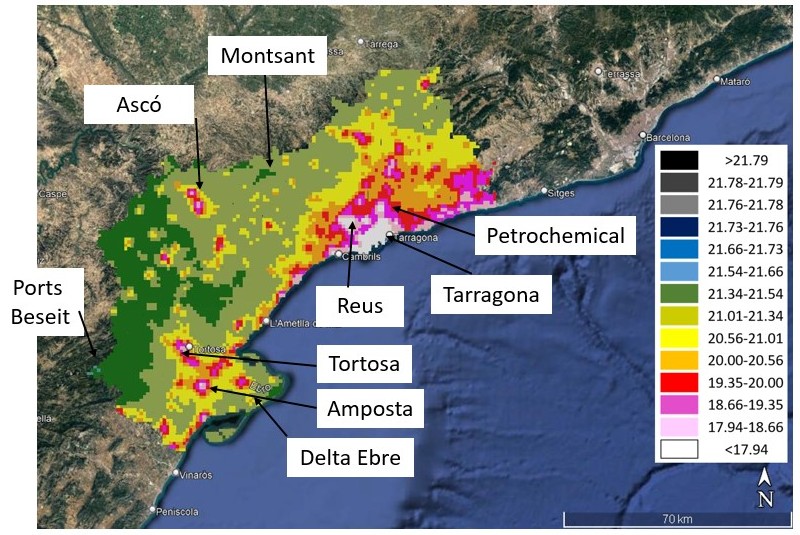}
\caption{ZSB maps in V band of Tarragona.}
\label{Tarragona_V}
\end{center}
\end{figure}

The resulting maps of the province of Tarragona with a sampling of 1x1 km are shown in figures \ref{tarragona_B} and \ref{Tarragona_V} for the B and V bands respectively.

Although the general pattern and values of the same locations coincide with the maps presented in the previous section, the higher number of sampled location provides a much greater detail of sky brightness patterns and sources. 

Both maps show bright values of sky brightness within and around Tarragona's metropolitan area, with several locations well below 20 and 18 mag/arcsec\textsuperscript{2} for the B and V bands respectively. The brightest pixel is the same for both bands and it is located in the port of Tarragona with a value of 15.73 mag/arcsec\textsuperscript{2} in V and 17.18 mag/arcsec\textsuperscript{2} in B.

There are three main sources easily distinguishable: Tarragona, Reus and the petrochemical complex. The light pollution produced by them extends several kilometers, affecting the interior areas of the region. 

With this sampling resolution it is easier to spot and analyse secondary sources. First, there is a narrow area that follows the coast north of Tarragona and in a lesser brightness also to the south. Second, middle sized cities as Amposta and Tortosa, located at the gates of Ebre's delta; or even smaller ones located in the south western border. Finally, special cases as the nuclear plant placed in the small city of Asc\'o, located 50 km north from Tortosa. 

The darkests areas, that reach 21.5 mag/arcsec\textsuperscript{2} for V and 22.5 mag/arcsec\textsuperscript{2} for B, are mainly in the Ports de Beseit Natural Park, small nuclei close to Montsant and in the eastern part of Ebre's delta.

\subsection{Comparison between B and V bands}

\begin{figure}[h!]
\begin{center}
\includegraphics[width=13cm]{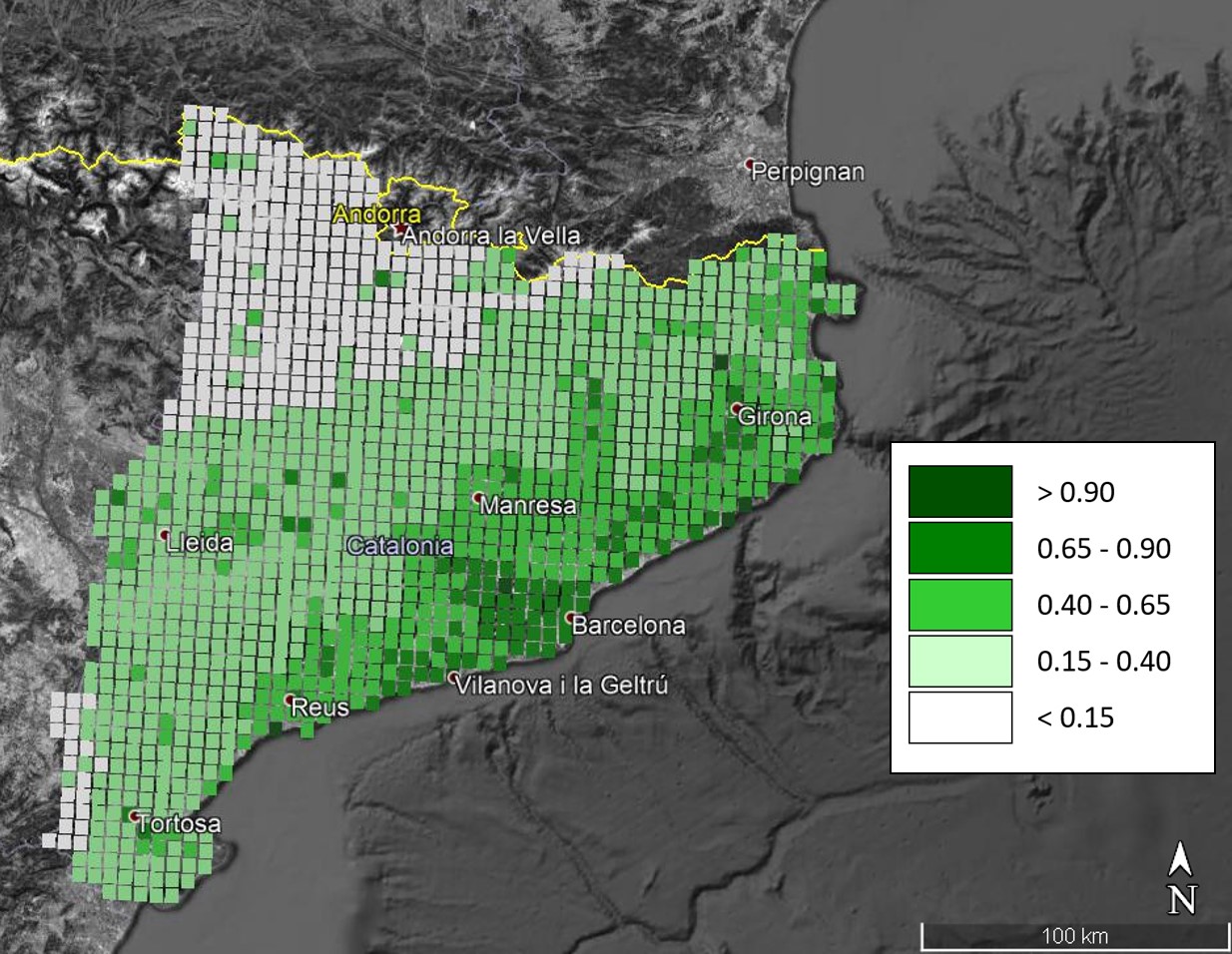}
\includegraphics[width=11cm]{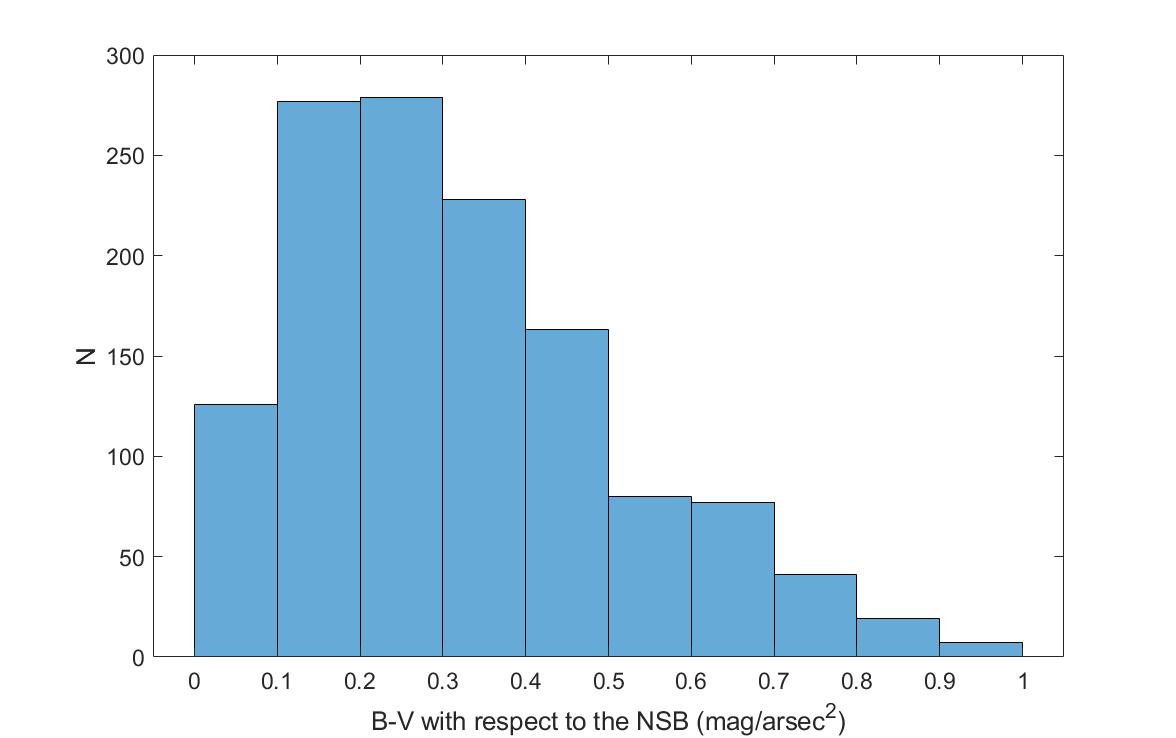}
\caption{Comparison between V and B (B-V) bands sky brightening with respect the natural brightness. Top: Catalonia map. Bottom: histogram.}
\label{bvmap}
\end{center}
\end{figure}

Comparing the results for both bands for the whole Catalonia region, we derive that the sky is more polluted in the V band than in the B band. Mainly because most of the artificial light sources have an emission more dominant in the V band as well. We created a comparison map (see figure \ref{bvmap}) between the two bands with respect to the natural brightness level of each band using the colour excess astronomic concept:

\begin{equation}
\label{appexcesscolor}
E_{B-V}=(B-V)_{total}-(B-V)_{natural}
\end{equation}

Positive values indicates a sky with more V band light proportion than the natural value. Negative values, a sky with more B band light proportion than the natural value.

There is not a single location where the brightening in the B band is greater than in the V band. The locations that have a more similar brightening in both bands are the less polluted ones (NW and SW) as the natural brightness is dominant. The locations that have a bigger difference follow the same pattern, they are the most polluted ones in general. However, there are three cases that do not follow that pattern (they are bright but not significantly brighter in V compared to B) due to their lighting system is rich in blue light (see table \ref{inventory_zbm}): Tarragona metropolitan area, Lleida and Andorra. 

\subsubsection{Comparison to Falchi's map}
\label{comparison_falchi1}
\begin{figure}[h!]
\begin{center}
\includegraphics[width=15cm]{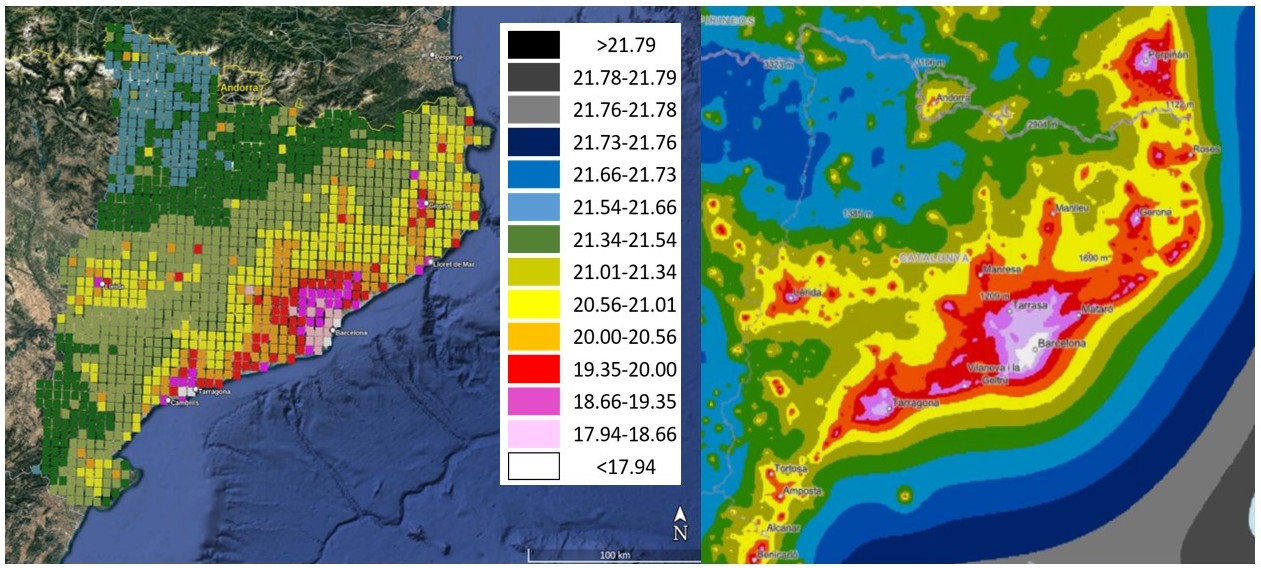}
\caption{ZSB maps in V band of Catalonia. Left: Illumina map. Right: \citep{falchiatlas}. }
\label{ZBMFalchiV2016}
\end{center}
\end{figure}

We proceeded to compare the ZSB maps of Catalonia to the one obtained by \citep{falchiatlas} for the V band (see figure \ref{ZBMFalchiV2016}).  In Falchi's map the natural brightness was set at 22.0 for the V band. We recomputed the brightness scale of that map using a natural brightness of 21.8 in order to have the same reference value than in our maps. 

Both maps assess a similar night sky (3 years apart) but the spatial sampling is not the same: 1x1 km in Falchi's map and 5x5 km in the map we created. A higher sampling creates a less patchy map with smoother transitions. We choose to compare first the map of the whole region of Catalonia as we are interested in the general trends of both methodologies. Using Catalonia map instead of Tarragona's provides a better test case as Barcelona dominates the light pollution situation in the region. 

Both maps show an almost equal brightness pattern. Apart from the general traits described in the previous section (metropolitan areas, dark values NW and SW, etc.), both map show light pollution trails from Barcelona towards secondary sources as Girona, Vic, Manresa and Igualada. 

On the other hand there are two big differences noticeable at first sight. First, the area of influence of big light pollution sources (middle-source too but in a lesser extent) is widely spread in Falchi's map. Second, darkness gradient once sky brightness reaches 21.0 is greater in Falchi's map, see the difference of the thickness of both light green (21.01-21.34) and dark green (21.34-21.54) between the two maps. The main hypothesis for why we see this difference are orographic blocking effect, atmospheric content and upward emission function. Falchi's map does not take into account the blocking effect of the orography and the obstacles. For instance, Barcelona is surrounded in the north by Collserola mountain range. Tarragona metropolitan area too is surrounded in the NW by Prades Hills. Moreover, different atmospheric characterization could contribute to the trend observed as a higher content in aerosols increases the sky brightness at short distance from the source but decreases further away. Another contributing factor to this difference could be the upward emission function definition. In Falchi's work the upward emission is dominated by a Lambertian (complemented with two other emission functions, one with a peak in very low angles and the other one with a peak at intermediate angles). On the other hand, Illumina v2 uses specific information of the obstacle and luminaries characteristics to define and ad-hoc emission function for each source.

Regarding dark areas, the darkest one in both maps is in the NW, they agree on  sky brightness values between 21.54 and 21.66 mag/arcsec\textsuperscript{2} although in Falchi's map there is a reduced number of locations between 21.66 and 21.73 mag/arcsec\textsuperscript{2}. In the SW the sky is also very dark, in the farthest corner reaching levels between 21.54 and 21.66 mag/arcsec\textsuperscript{2}. 

The difference in spatial resolution complicates the comparison of brightness values of middle-size sources, as the nearest estimated location could be outside its borders. For the cases where the studied location is within its borders, such as Vielha, La Seu d'Urgell, Tremp, La Pobla de Segur, Tortosa, Valls, Banyoles and Solsona, both maps agree, but in general we see differences up to 0.4 mag/arcsec\textsuperscript{2}, being darker in the Illumina map. For instance Manresa, Figueres, Roses, Mollerusa, T\`{a}rrega, Flix and Amposta. 

Barcelona is darker in the Illumina map, 18.2-18.4 mag/arcsec\textsuperscript{2}, than in Falchi's map, below 18.0 mag/arcsec\textsuperscript{2}. This difference is not seen in other big sources as Tarragona, Girona and Lleida, where the values of the locations closer to the centre of the source show the same value in both maps.

We proceeded to use the 1x1km map of Tarragona for the comparison. The increased resolution of Tarragona map, that matches the one from figure \ref{ZBMFalchiV2016}, makes it easier to test the effect of medium and small size light pollution sources. Location and values for these kind of sources agree much better in this map than in the one with less resolution. However, the main differences between the two methodologies regarding the spread of sky brightness around cities and towns continues to differ. This confirms what was explained previously in this section.

\subsection{Comparison to SQM dynamic measurements}
\label{sqm_measurements}
Extensive dynamic SQM measurement in Tarragona province were made in 2018 and 2019 \citep{linares_thesis}. The measurement campaign consisted in installing a SQM on the top of a vehicle and cover the roads and tracks of different regions. Each road is normally covered only once with the aim to assess as much territory as possible, but sometimes due to the road plan they are covered twice. All measurements were taken in moonless and cloudless nights. Those measurements, after being processed and filtered, can be used to create sky brightness cell maps of the paths covered (for more information about the filtering and processing of the data see \citep{Ribas2016,linares_thesis}). The SQM-LU used for this campaign is only used during the night, therefore it is not affected by the degradation observed in permanently installed devices due to exposure to the sun \citep{barasqm4}.

\begin{figure}[h!]
\begin{center}
\includegraphics[width=11cm]{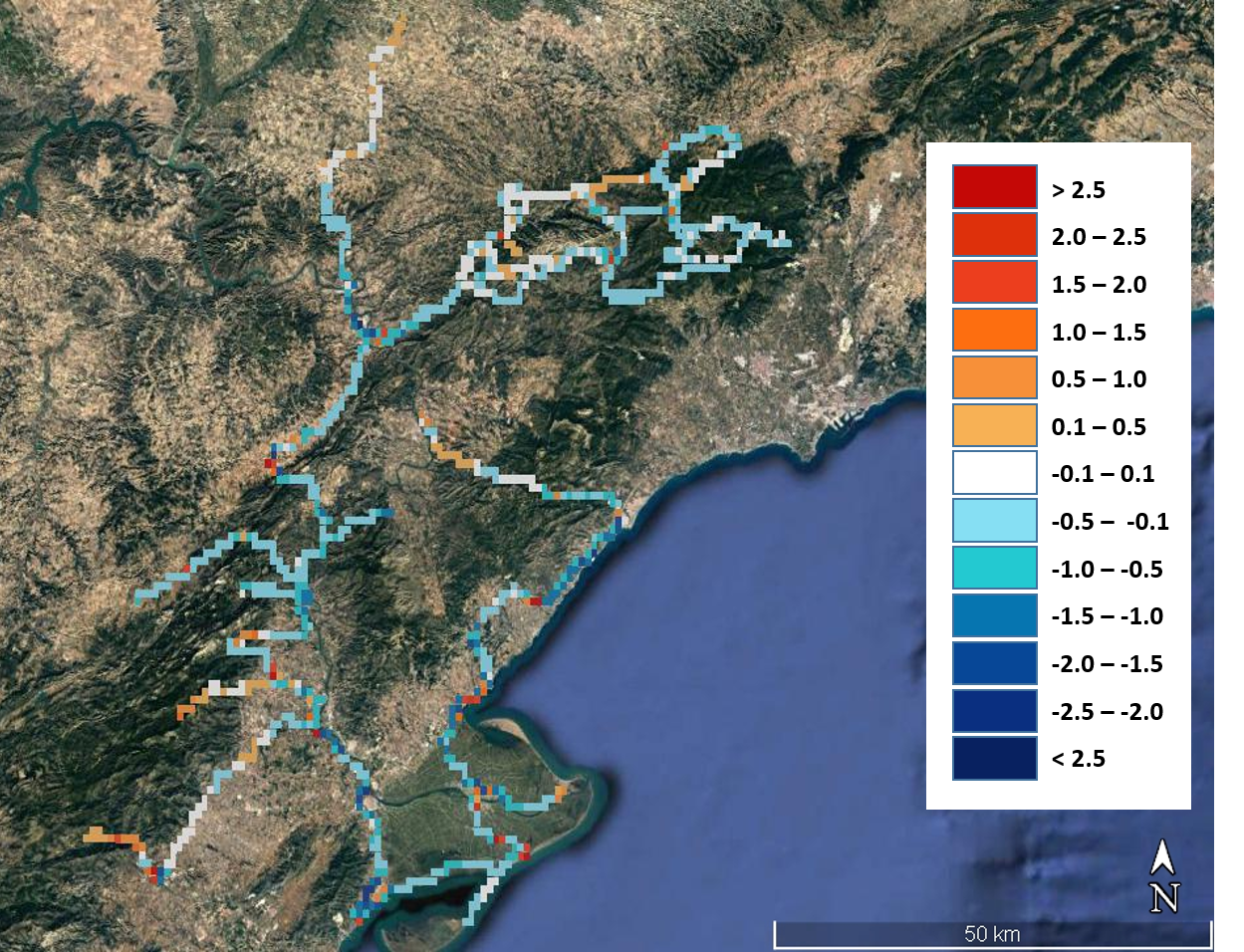}
\caption{Sky brightness difference map between the ZSB map and the SQM map.}
\label{diff_sqm}
\end{center}
\end{figure}

\begin{figure}[h!]
\begin{center}
\includegraphics[width=14cm]{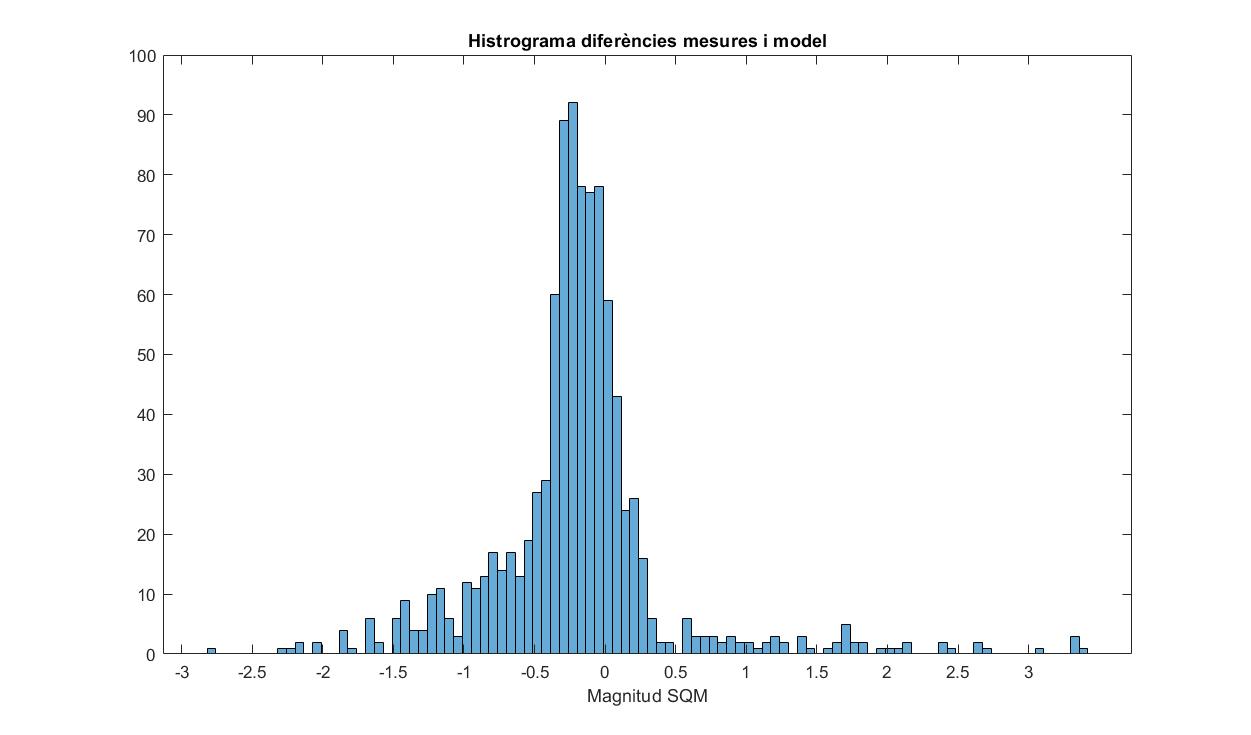}
\caption{Histogram of the difference between the ZSB map and the SQM map.}
\label{histo_sqm}
\end{center}
\end{figure}

Figure \ref{diff_sqm} shows the difference between the maps created from measurements and the closest location where the ZSB has been computed. Positive values mean that measurements are brighter and negative values that measurements are darker. In general, there is a slight tendency for computed values to be brighter (see figure \ref{histo_sqm}). In this paper the aim was to present the difference in a qualitative analysis as a proper comparison between these two methodologies is very complex, it includes many variables of unknown outcome. First, the two maps could not be assessing exactly the same location, a difference of 500 m near the border of villages could be enough for a significant change in brightness. Second, SQM measures are sensitive to different atmospheric conditions whereas computations were done with constant values. Third, the natural sky brightness at zenith varies along the year, mainly because of the presence of the Milky Way. In following projects a natural sky model, GAMBONS \citep{Gambons}, will be used to remove that variable. 

\section{Conclusion}

The methodology presented in this paper has been proven as valid and accurate for assessing the sky brightness over large areas. In this first approach we used it to create ZSB maps in V and B bands with different sampling resolution. However, Illumina (the light pollution model the methodology is based on) allows to play with the variables. As such, it is possible to create maps that show other sky brightness indicators, other areas, sampling resolution and also other bands. Additionally, it can be used as a prediction tool for studying the sky brightness in hypothetical situations.

The downside of our approach is the time needed to create such maps. The maps of Catalonia took around 50.000 CPU hours and the ones of Tarragona, 200.000 CPU hours. However, in the near future this number will be drastically reduced with the new versions of the Illumina model. Particularly, the simplified version of the model called Illumina Light will fasten up the computation by at least an order of magnitude due to an optimized division of the spectral range to cover the V band and the geographical domain for each point.

The methodology of Sánchez de Miguel \citep{sanchez2019} allows to define more accurate inventories. However, it is important to keep working with public inventories too. Having both will allow estimating the percentage of public and private light that a city is emitting. This information is useful to assess the real impact of changing the public lighting system.

Our approach assumed some simplifications that should be tested in future studies to see how important is the improvement in precision of the results. First, in this study the natural light is assumed as constant, natural light models such as GAMBONS \citep{Gambons} could be used instead. Second, Illumina assumes the same atmospheric content for the whole geographic domain, that could introduce some error, particularly when distant sources are affecting the sky over a location. Third, the model also assumes a plane-parallel atmosphere for estimating the interactions between photons and atmospheric particles (the curvature of the Earth is accounted for the position and angles of the light sources as well as the the blocking effect of the horizon and obstacles). Finally, Illumina allows a more refined definition of the inventory that the one used in this study. Cities and towns could be divided in several regions, and it is even possible to define lamp lights individually. This level of precision was out of the scope for this first methodology validation study.

A sampling resolution of 1x1 km is required to study fine details such as as the impact of middle and small towns in their surroundings. Our findings confirms what was stated in \citep{Barasampling}. On the other hand, due to the CPU time requirements, a resolution of 5x5  km could be interesting for studies that only need to point out the main sources of light pollution and the general pattern of sky brightness in the area studied.

Regarding the sky brightness over the areas studied, we observed that Catalonia has a complex pattern of light pollution. It is due to the heterogeneity of the country regarding population distribution and the orography of the territory. Additionally, we found out that light pollution is dominant in the V band with respect to the B band. The main traits of the sky brightness over Catalonia are:
\begin{itemize}
    \item Barcelona and Tarragona metropolitan areas are by far the greatest and second greatest artificial light sources of Catalonia. Zenith sky brightness in a radius of tens of kilometres are heavily influenced by them.
    \item In a lesser extent, the rest of the coastal line is also a source of light pollution. There is a general pattern of darkening as we move away from the coast. The exception is Delta de l'Ebre that is a protected area and barely populated.
    \item The most polluted skies apart from the ones already mentioned are over Lleida and Girona. 
    \item The darkest regions are the western Pyrenees, Montsec mountain range and its surroundings, and Ports de Beseit NP.
\end{itemize}

The ZSB map for the V band was compared to the reference data in this field, the light pollution atlas map of Falchi et al \citep{falchiatlas}. Both maps show an almost equal brightness pattern. The biggest difference between the two methods is the area of influence of big light pollution sources. The differences could be explained by the different methodologies used, in particular the accounting for orography and obstacles, and the emission function. However, a quantitative study should be performed to discard other variables that could also play a role. 

The maps presented in high resolution have been qualitatively compared to maps created from SQM measurement. There is a slight shift towards the measurements being darker. The cause of this difference is not obvious as it could be explained by diverse factors: atmospheric conditions, natural light, not perfect matches of the locations mapped, and private lighting (which vary rapidly and is difficult to characterize).

\section*{Acknowledgements}

We applied the sequence-determines-credit approach for the sequence of authors \citep{tscharntke2007author}. 
Hector Linares Arroyo thanks the Fonds de recherche du Québec - Nature et technologies along with the Canadian Space Agency and the Generalitat de Catalunya for supporting this research. M. Aubé thanks Fonds de recherche du Québec - Nature et technologies for their support. 
Part of the  high performance computing resources required for this work was provided by Calcul Québec and Compute Canada through M. Aubé's affiliations to these organisms. 
Generalitat de Catalunya has funded the map of Tarragona and the evaluation of results, as well as has provided information on SQM, SQC and ASTMON measurements, and the inventory of some cities.

\section*{}
\bibliographystyle{elsarticle-num-names}
\bibliography{biblio}

\appendix
\section{Comparion between SQM and V filters}
\label{appendix}

The aim of this appendix is to prove numerically the that SQM readings in the presence of artificial light are normally brighter than readings in the V band of the Johnson system.

The approach consists in filtering the emission of the three main technology lamps present in the territory (HPS, LED 3000K and LED 4000K) by the two filters and compare them to their 0 point.

$$R=(T_{sqm}/ref_{sqm})/(T_V/ref_V)$$

Being R, the ratio between the two readings in terms of radiance; $T_{sqm}$ the emission of the lamp filtered by the SQM filter; $ref_{sqm}$ the reference (or 0 point) for the SQM; $T_V$ the emission of the lamp filtered by the V filter; and $ref_V$ the reference (or 0 point) for the V filter.
The equation implies that if R is greater than 1 the reading for the SQM filter is brighter than in the V filter.

Both filters must be normalized in the same manner that they were when computing the 0 point. For the V filter \citep{Bessell_1979} this was done with their maximum transparency equal to 1. For the case of SQM it is implicitly stated that the same normalization was used but in the web-page of the company all the plots of the spectral response show a maximum of 0.9. In this case we used the worst case scenario to prove our point, i.e. using 0.9 as a maximum.

For the V filter the zero point corresponds to 140.4 W/m$^2$/sr \citep{Bessell_1979} and for the SQM 131.4 W/m$^2$/sr (source Unihedron web-page, particularly it is stated as 108000 cd/m$^2$, with a conversion factor of $1.3x10^{-3}$ to obtain W/m$^2$/sr).

To ensure an acceptable level of precision we are working with a spectral resolution of 1 nm.

\begin{table}[h]
\begin{center}
\begin{tabular}{|p{4cm}|p{3cm}|}
\hline
Technology lamp  & R \\
\hline
High Pressure Sodium  & 1.10 \\
LED 3000K  & 1.15 \\
LED 4000K & 1.25 \\
\hline
\end{tabular}
\caption{R defined as the ratio between the brightness over the 0 point of the filter SQM and V of the emission spectrum of the most typical lights in Catalonia.}
\label{table_R}
\end{center}
\end{table}

\begin{figure}[h!]
\begin{center}
\includegraphics[width=14cm]{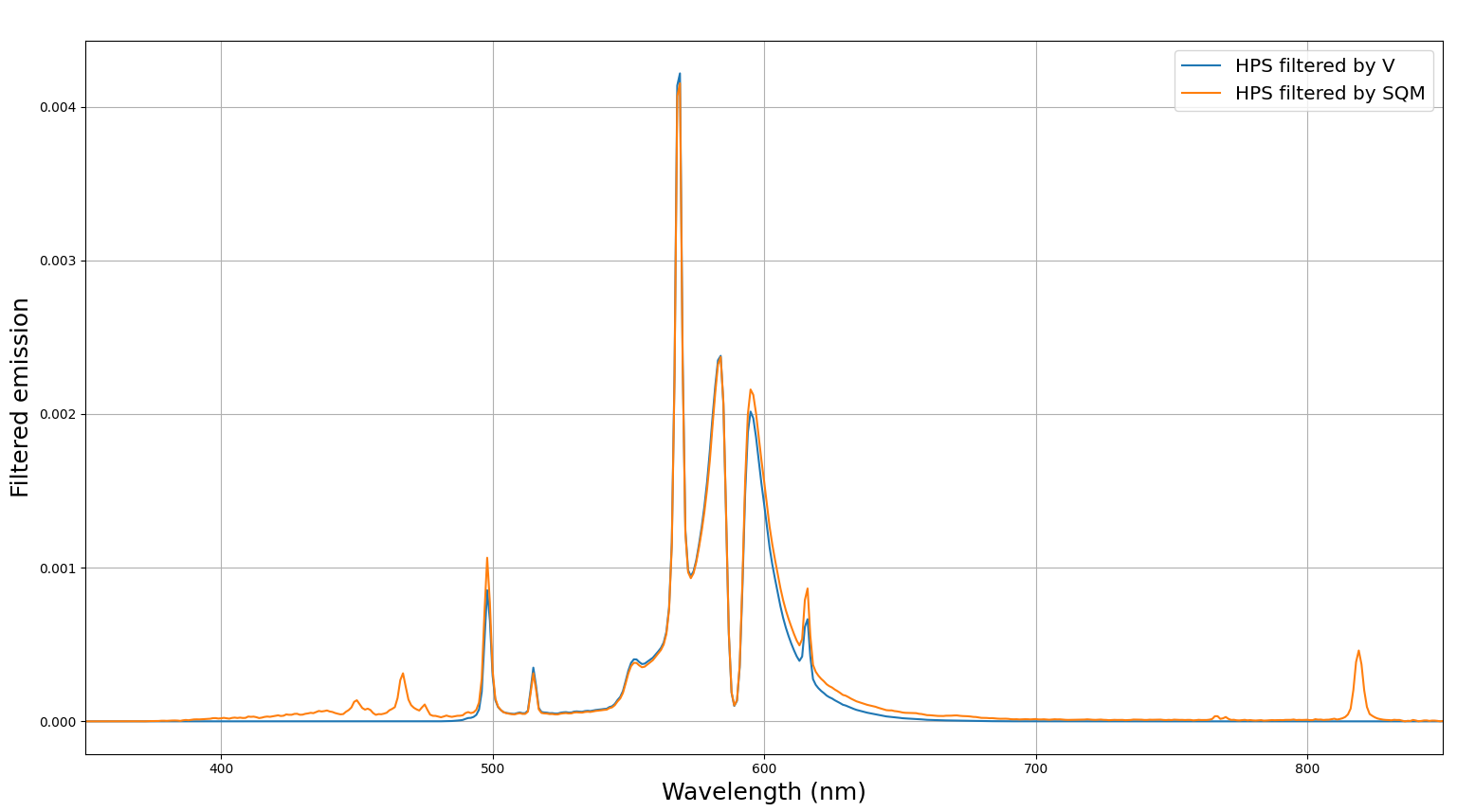}
\caption{HPS filtered emission. In orange SQM filter and in blue V filter.}
\label{ratio_hps}
\end{center}
\end{figure}

\begin{figure}[h!]
\begin{center}
\includegraphics[width=14cm]{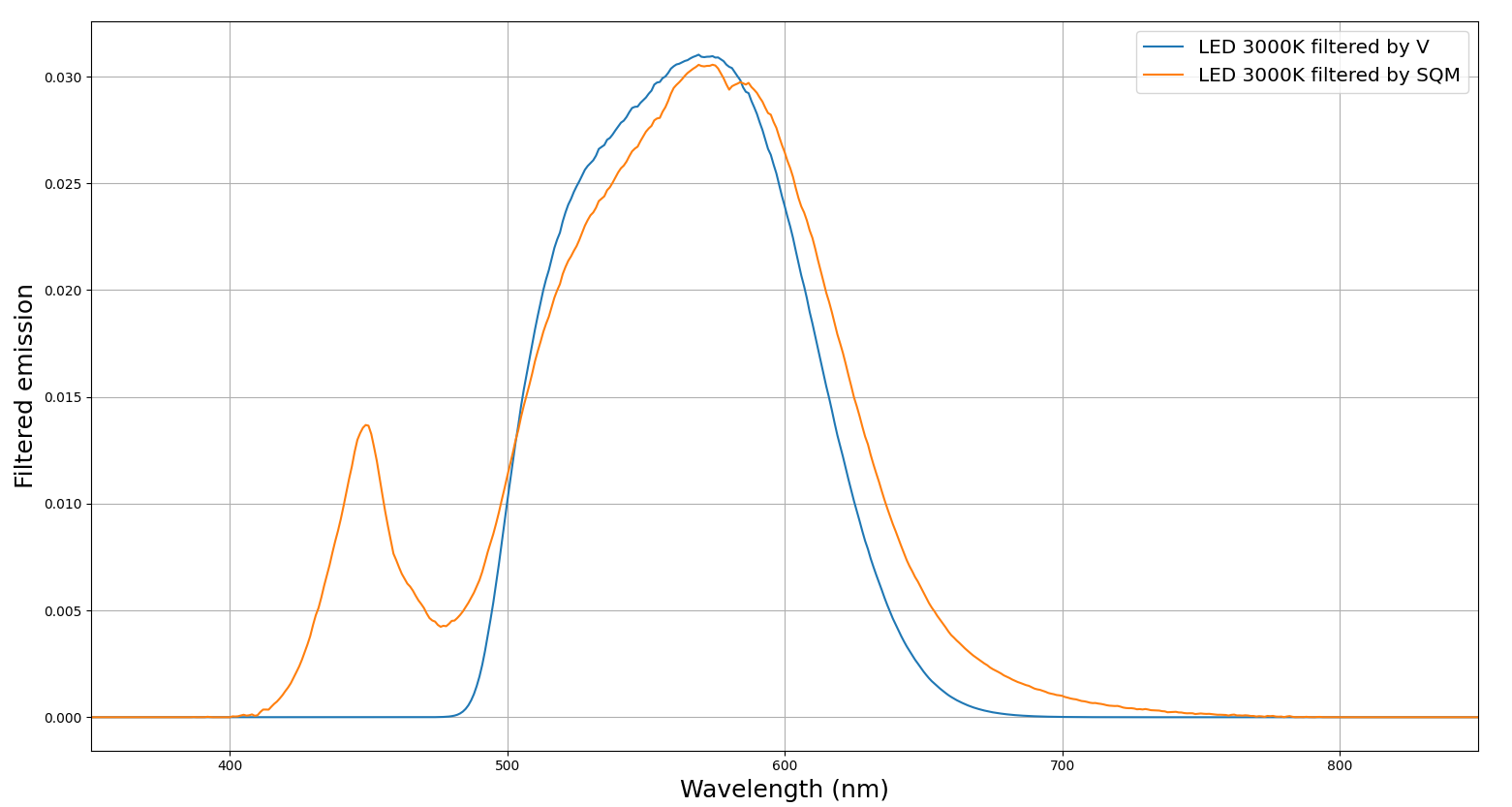}
\caption{LED 3000K filtered emission. In orange SQM filter and in blue V filter.}
\label{ratio_3led}
\end{center}
\end{figure}

\begin{figure}[h!]
\begin{center}
\includegraphics[width=14cm]{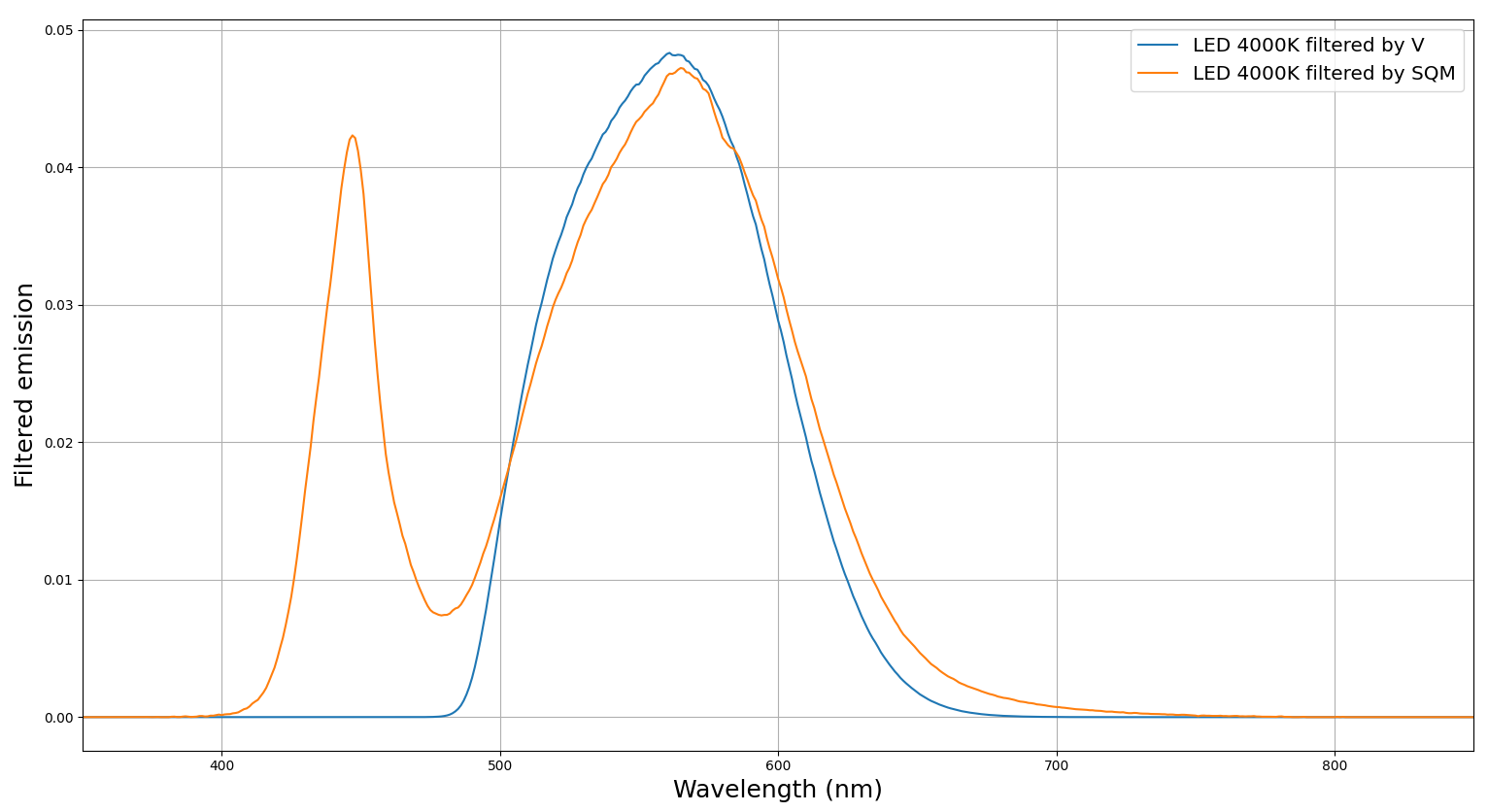}
\caption{LED 4000k filtered emission. In orange SQM filter and in blue V filter.}
\label{ratio_4led}
\end{center}
\end{figure}

As seen in table \ref{table_R} the ratio is over 1 in the three cases. That means that the SQM reading, in the presence of only artificial light, is 10\% brighter for HPS, 15\% for LED of 3000K and 25\% for the LED of 4000K. The difference between the emission spectrum filtered by each filter is shown in figures \ref{ratio_hps}, \ref{ratio_3led} and \ref{ratio_4led}.

In terms of magnitude depending on the ratio between the different technologies it means between 0.1 and 0.3 mag/arcsec\textsuperscript{2} brighter for the SQM. The difference is expected to decrease as the natural light increases its weight on the total sky brightness. That is aligned with the trend we saw in figure \ref{chartzbm}, where the biggest difference where seen in the most polluted locations (artificial radiance dominant), and the smallest in the most pristine locations. 

\end{document}